\documentclass[twocolumn,trackchanges]{aastex62}
\def\beq{\begin{equation}}
\def\enq{\end{equation}}  
\def\ba{\begin{eqnarray}}
\def\ea{\end{eqnarray}}
\def\<{\langle}
\def\>{\rangle}

\def\eps{\epsilon}

\usepackage{amsmath}
\usepackage{graphicx}
\usepackage{amsfonts}
\usepackage{natbib}
\usepackage{color}
\usepackage{epstopdf}
\usepackage{graphicx}
\usepackage{multirow}
\usepackage{longtable}
\usepackage{rotating}
\usepackage{calrsfs,euscript,mathrsfs,amssymb}

\shorttitle{Using EGB to constrain $H_0$ and $\Omega_{\rm m}$}
\shortauthors{Zeng \& Yan}

\begin{document}

\title{Using the Extragalactic Gamma-Ray Background to Constrain the Hubble Constant and Matter Density of the Universe}

\author{Houdun Zeng}
\affil{Key Laboratory of Dark Matter and Space Astronomy, Purple Mountain Observatory, Chinese Academy of Sciences,
Nanjing 210008, China; zhd@pmo.ac.cn}

\author{Dahai Yan}
\affil{Key Laboratory for the Structure and Evolution of Celestial Objects, Yunnan Observatories, Chinese Academy of Sciences, Kunming 650216, China; yandahai@ynao.ac.cn}
\affil{Center for Astronomical Mega-Science, Chinese Academy of Sciences, 20A Datun Road, Chaoyang District, Beijing 100012, China}
\affil{Department of Astronomy, Key Laboratory of Astroparticle Physics of Yunnan Province, Yunnan University, Kunming 650091, China}







\begin{abstract}
The attenuation produced by extragalactic background light (EBL) in $\gamma$-ray spectra of blazars 
has been used to constrain the Hubble constant ($H_0$) and matter density ($\Omega_{\rm m}$) of the Universe.
We propose to estimate $H_0$ and $\Omega_{\rm m}$ using the well measured $>$10 GeV extragalactic $\gamma$-ray background (EGB).
This suggestion is based on the facts that the $>$10 GeV EGB is totally explained by the emissions from blazars, and 
an EBL-absorption cutoff occurs at $\sim$50 GeV in the EGB spectrum.
We fit the $>$10 GeV EGB data with modeled EGB spectrum.
This results in $H_0=64.9^{+4.6}_{-4.3}\rm \ km\ s^{-1}\ Mpc^{-1}$ and $\Omega_{\rm m}=0.31^{+0.13}_{-0.14}$.
Note that the uncertainties may be underestimated due to the limit of our realization for EBL model.
 $H_0$ and $\Omega_{\rm m}$ are degenerate in our method.
 Independent determination of $\Omega_{\rm m}$ by other methods would improve the constraint on $H_0$.
\end{abstract}



\section{Introduction} \label{sec:intro}

A precise and accurate measurement of the Hubble constant ($H_0$) would provide deep understanding of fundamental physics questions.
Multiple paths to independent estimates of $H_0$ are needed in order to access and control its systematic uncertainties \citep{Suyu}.

Gamma-ray astronomy provides a new approach to estimate $H_0$ \citep{Salamon,Mann}.
The optical depth of the $\gamma$-ray photons emitted by extragalactic objects, $\tau_{\gamma\gamma}$, scales as $n_{\rm EBL}\sigma_{\rm T}l$, 
where $n_{\rm EBL}$ is the photon density of the extragalactic background light (EBL), 
$\sigma_{\rm T}$ is the Thomson cross section, and $l$ is the distance from the $\gamma$-ray source to Earth.
$l$ is inversely proportional to $H_0$, 
and $n_{\rm EBL}$ also depends on $H_0$. 
Therefore, through determining the optical depth $\tau_{\gamma\gamma}$,  one can estimate $H_0$.

Such an approach has been pursued by latter studies.
With simulated TeV spectra of blazars, \citet{Blanch} studied the possibility of using $\gamma$-ray absorption to constrain cosmological parameters.
Using the EBL density based on galaxy counts, \cite{Barrau} derived $H_0>74\rm \ km\ s^{-1}\ Mpc^{-1}$ at the 68\% confidence level, from the TeV spectrum of Mrk 501.
With the cosmic $\gamma$-ray horizon extracted from multiwavelength observations of TeV blazars \citep{Dom13a},
\citet{Dom13} derived  $H_0=71.8_{-5.6}^{+4.6}\rm\ (stat)^{+7.2}_{-13.8}\ (syst) \ km\ s^{-1}\ Mpc^{-1}$.
\citet{Biteau} derived $H_0=88\pm13\rm\ (stat)\pm13\ (syst) \ km\ s^{-1}\ Mpc^{-1}$ by analyzing 106 TeV spectra of 38 blazars.
The {\it Fermi} Large Area Telescope ({\it Fermi}-LAT) observations of blazars provide good determinations of $\tau_{\gamma\gamma}$ \citep{FermiSci}.
Using $\tau_{\gamma\gamma}$ measured from {\it Fermi}-LAT GeV spectra \citep{FermiSci} and TeV spectra \citep{Desai}, 
\citet{Dom19} derived $H_0=68.0_{-4.1}^{+4.2}\rm \ km\ s^{-1}\ Mpc^{-1}$ and $\Omega_{\rm m}=0.17^{+0.07}_{-0.08}$ with the combination of the EBL models of \citet{Finke10} and \citet{Dom}.
The constraint on $H_0$ from $\gamma$-ray attenuation has been significantly improved in the past ten years.

The above constraints on $H_0$ are all derived from point sources.
Here, we propose to constrain $H_0$ and $\Omega_{\rm m}$ using the extragalactic $\gamma$-ray background (EGB).
The EGB spectrum has been well measured from 0.1 GeV to $\sim$800 GeV by the {\it Fermi}-LAT.
This spectrum can be described by a power law with a photon index of 2.32 that is exponentially cut off at $\sim$50 GeV \citep{Ackermann}.
The cutoff is caused by the EBL absorption \citep{Ajello}.
Similar to the idea proposed by \citet{Salamon}, the $\gamma$-ray absorption in the EGB spectrum could also be used to constrain the cosmological parameters.

EGB is dominated by the emission of $\gamma$-ray blazars \citep{Ajello,Ackermann16}.
With the source count distribution of hard-spectrum blazars, \citet{Ackermann16} estimated that 
blazars can explain almost the totality ($86^{+16}_{-14}\%$) of the $>$50 GeV EGB.
In particular, the calculation performed with improved luminosity function (LF) and modeling of the
spectral energy distributions (SEDs) of blazars showed that blazars account for the totality of the $\geq$10 GeV EGB \citep{Ajello}.
Besides, modeling of the EGB spectrum also depends on $H_0$.
Therefore, we can use the above information to constrain $H_0$ and $\Omega_{\rm m}$.

\section{Method}

\subsection{Calculation of the EGB spectrum}

We follow \citet{Ajello} to compute the EGB spectrum contributed by blazars,
\begin{equation}
\begin{split}
F_{\rm EGB} (E_{\gamma}) = \int\limits^{\Gamma_{\rm max}=3.5}\limits_{\Gamma_{\rm min}=1.0}{\rm d}\Gamma \int\limits^{z_{\rm max}=6}\limits_{z_{\rm min}=10^{-3}}{\rm d}z\\
\times\int\limits^{L_{\gamma}^{\rm max}=10^{52}}\limits_{L_{\gamma}^{\rm min}=10^{43}}{\rm d}L_{\gamma} \cdot \Phi(L_{\gamma},z,\Gamma)
\cdot \frac{dN_{\gamma}}{dE}\cdot \frac{dV}{dzd\Omega}\\
\times[{\rm ph\ cm^{-2} s^{-1} sr^{-1} GeV^{-1}}],
\label{eq:egb}
\end{split}
\end{equation}
where the LF, $\Phi(L_{\gamma},z,\Gamma)$ (at redshift $z$, for sources of $\gamma$-ray luminosity $L_{\gamma}$), 
is described as a broken power law multiplied by the photon index distribution $\frac{dN}{d\Gamma}$ \citep[Equation~(1) in][]{Ajello}.
The $\gamma$-ray spectrum of each blazar, $\frac{dN_{\gamma}}{dE}$, is modeled as a broken power law \citep[Equation~(11) in][]{Ajello}.
$\frac{dV}{dzd\Omega}$ is the comoving volume
element per unit redshift and unit solid angle, which is written as,
\begin{equation}
\frac{dV}{dzd\Omega}=\frac{cd^2_{L}}{H_0(1+z)^2}\frac{1}{E(z)},
\end{equation}
where $E(z)=[\Omega_{\Lambda}+\Omega_{\rm m}(1+z)^3]^{1/2}$, $\Omega_{\Lambda}=1-\Omega_{\rm m}$ in a flat $\Lambda$CDM cosmology, and $d_{L}$ is the luminosity distance.

\subsection{Absorption of $\gamma$-rays}

The optical depth of the $\gamma$-ray photons emitted at redshift $z$ as a function of observed $\gamma$-ray photon energy, 
$E_{\gamma}$, is calculated by \citep[e.g.,][]{Razzaque}

\begin{eqnarray}
\tau_{\gamma\gamma} (E_{\gamma}, z) &=& 
c\pi r_e^2 \frac{m_e^4c^8}{E_{\gamma}^2}
\int_0^z \frac{dz_1}{(1+z_1)^2} \left| \frac{dt}{dz_1} \right| 
\nonumber \\ && ~\times
\int_{\frac{m_e^2 c^4}{E_{\gamma}(1+z_1)}}^{\infty} {d\eps_1} 
\frac{\eps_1u_{\rm EBL}(\eps_1,z_1)}{\eps_1^4} {\bar \varphi} (s_0),
\label{gg_opacity}
\end{eqnarray}
where $\left| \frac{dt}{dz_1} \right| =\frac{1}{H_0(1+z_1)E(z_1)}$, $s_0=E_{\gamma}\eps_1(1+z_1)/m^2_ec^4$, and ${\bar \varphi} (s_0)$ is adopted from \citet{Gould}.
We use the model of \citet{Razzaque} to calculate the comoving
EBL density,

\ba
\eps u(\eps,z)= (1+z)^4\eps^2
{\cal N} \int_{z}^\infty dz'' 
\left| \frac{dt}{dz''} \right| \psi(z'')
\nonumber \\ \times 
\int_{M_{\rm min}}^{M_{\rm max}} dM \left( \frac{dN}{dM} \right)
\nonumber \\ \times 
\int_{z_{\rm d} (M,z')}^{z''} dz' 
\left| \frac{dt}{dz'} \right| f_{\rm esc} (\eps') 
\frac{dN(\eps',M)}{d\eps' dt} (1+z'),~
\label{final-photon}
\ea
where $\psi(z)$ is the star formation rate (SFR) in unit of $M_\odot$~yr$^{-1}$
~Mpc$^{-3}$,  $\frac{dN}{dM}$ is the initial mass function (IMF), $f_{\rm esc} (\eps)$ is the escape fraction
of photons from the host galaxy,
and $\frac{dN(\eps,M)}{d\eps dt}$ is the total number of photons emitted from a star.
The normalization is determined by ${\cal N}^{-1} =
\int_{M_{\rm min}}^{M_{\rm max}} dM (dN/dM)M$.
$z_{\rm d} (M,z)$ is the redshift of the star (born at redshift $z$) that had
evolved off the main sequence. See \citet{Razzaque} for more details.

The uncertainties in modeling the EBL density primarily come from SFR and IMF.
We adopt the {\it Models B} and {\it C} in \citet{Razzaque}.
Both models use the same SFR \citep{Cole,sfr}, but different IMFs.
 {\it Model B} uses Salpeter A IMF \citep{Salpeter}, and  {\it Model C} uses Baldry-Glazebrook IMF \citep{Baldry}.

Note that the EBL model of \citet{Razzaque} only includes the contribution from starlight.
This underestimates the EBL density below 1 eV \citep{Finke10}, and consequently underestimates $\tau_{\gamma\gamma}$ 
above 0.3/(1+z) TeV.

\subsection{Verification of our calculations}
\begin{figure*}
 \centering
     \includegraphics[scale=0.6]{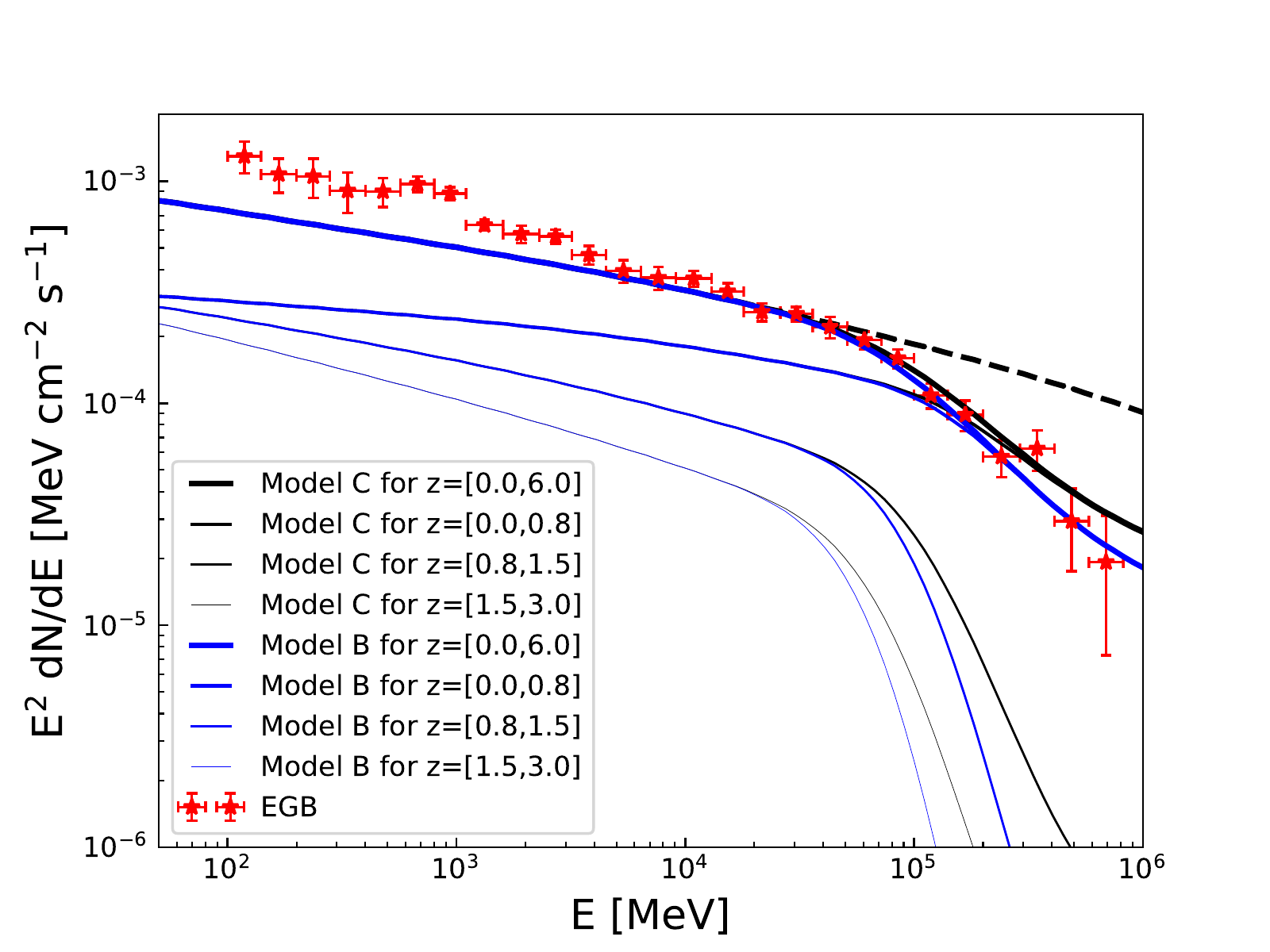}
\caption{Contribution to the EGB from blazars, calculated with the PLE LF in \citet{Ajello} and the EBL models in \citet{Razzaque}(solid lines). 
               The dashed line is the one without EBL absorption. 
                Data points are from \citet{Ackermann}.} 
\label{EGB}
\end{figure*}

We calculate the contribution to the EGB from blazars with the pure luminosity evolution (PLE)  LF in \citet{Ajello} 
and the EBL {\it Models B} and {\it C}. The parameters in Table~1 in \citet{Ajello} are used.
Here we adopt $H_0=67\rm \ km\ s^{-1}\ Mpc^{-1}$ and $\Omega_{\rm m}=1-\Omega_{\Lambda}=0.3$, same as that in \citet{Ajello}.
The results are shown in Fig.~\ref{EGB}.
One can see that EGB above 100 GeV can be explained by the emission from the blazars below the redshift of 0.8,
whereas EGB between 10 GeV and 100 GeV can be explained by the blazars below the redshift of 1.5.

We compare our results with \citet{Ajello} who adopted the EBL model of \citet{Finke10}.
We found that our results are almost the same as that in \citet{Ajello} (see their Fig.~3) below 300 GeV.
Above 300 GeV, the intensity that we calculated with EBL {\it Model C} is higher than the one in \citet{Ajello}.
This is due to our underestimation of the EBL intensity.
However, we note that above 300 GeV, the intensity in \citet{Ajello} agrees with ours within the errors of the data points.

The results in Fig.~\ref{EGB} show two points: (1) the emission from blazars could be used to explain the EGB above $\sim10$ GeV;
(2) the difference between EBL models of \citet{Razzaque} and \citet{Finke10} has little impact on explaining the origins of EGB.

\section{Results}

Calculations of LF and SFR depend on the measurements of $H_0$ and $\Omega_{\rm m}$.
\citet{Ajello} constructed the LF with $H_0=67\rm \ km\ s^{-1}\ Mpc^{-1}$ and $\Omega_{\rm m}=0.3$.
In our purpose, the LF should be modified with different cosmological parameters.
Therefore, the LF in Equation~(\ref{eq:egb}) is,
\begin{eqnarray}
&&\Phi(L_{\gamma},z,\Gamma)dL_{\gamma}dzd\Gamma \nonumber \\ 
&&=\Phi_{\rm Ajello15}(L_{\gamma},z',\Gamma')\frac{dV/dz/d\Omega}{dV'/dz'/d\Omega'}dL_{\gamma}dz'd\Gamma'\ .~
\end{eqnarray}

The SFR in Equation~(\ref{final-photon}) is modified as \citep[e.g.,][]{Dom19},
\begin{equation}
\psi(z)=\psi_{\rm HB06}(z')\frac{H_0E(z)}{H'_0E'(z)}.
\end{equation}

The primed quantities are computed with $H'_0=67\rm \ km\ s^{-1}\ Mpc^{-1}$ for the LF,  and $H'_0=70\rm \ km\ s^{-1}\ Mpc^{-1}$ for the SFR, and $\Omega'_{\rm m}=0.3$.

\subsection{Dependence on $H_0$}

\begin{figure*}
 \centering
     \includegraphics[scale=0.5]{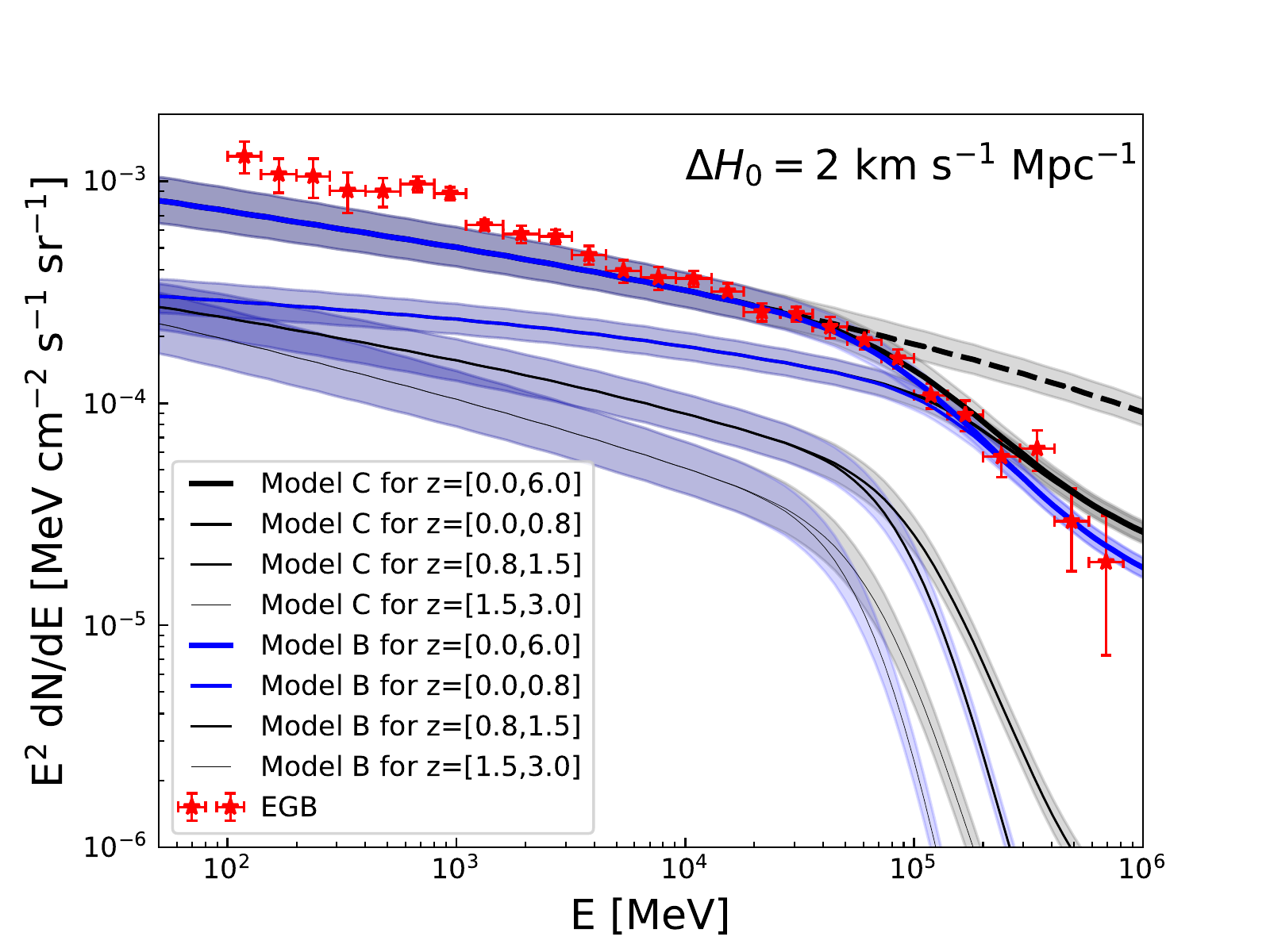}
     \includegraphics[scale=0.5]{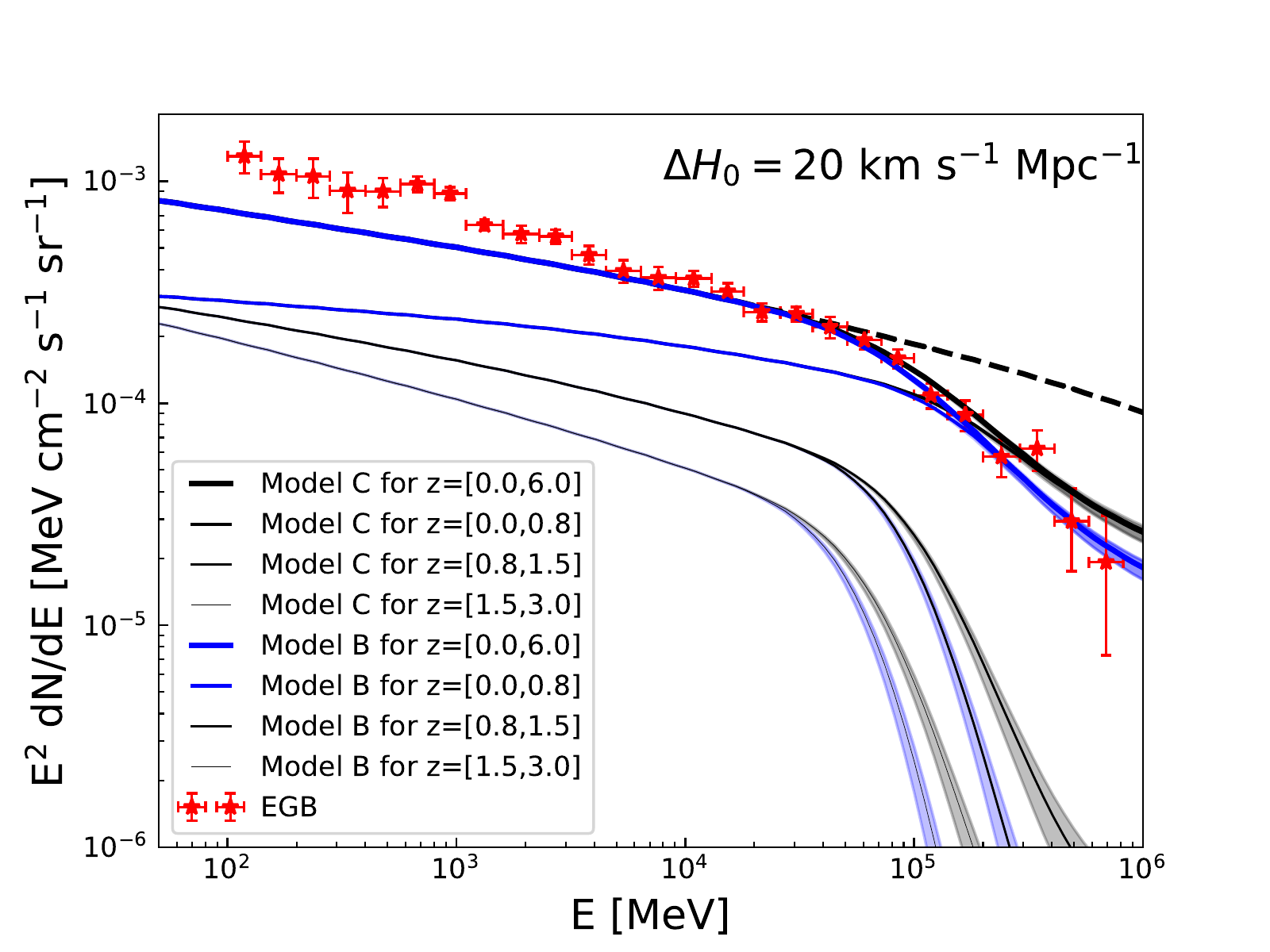}
\caption{Dependence on $H_0$. The results are produced by fixing $\Omega_{\rm m}=0.3$.
             {\it Left:} $H_0$ is varied from $65 \rm \ km\ s^{-1}\ Mpc^{-1}$ to $69 \rm \ km\ s^{-1}\ Mpc^{-1}$ in the calculation of the intrinsic EGB spectrum.
             {\it Right:} $H_0$ is varied from $47 \rm \ km\ s^{-1}\ Mpc^{-1}$ to $87 \rm \ km\ s^{-1}\ Mpc^{-1}$ in the calculation of $\tau_{\gamma\gamma} (E, z)$.  } 
\label{Dependence}
\end{figure*}

Calculations of both the intrinsic EGB spectrum and $\tau_{\gamma\gamma} (E, z)$ depend on $H_0$ and $\Omega_{\rm m}$.
In Fig.~\ref{Dependence}, we can see that the intrinsic spectrum strongly relies on $H_0$, 
especially at the energies below 100 GeV (left panel; $H_0$ is fixed to $67\rm \ km\ s^{-1}\ Mpc^{-1}$ in the calculation of the optical depth);
and the dependence of $\tau_{\gamma\gamma} (E, z)$ on $H_0$ occurs at the energies above 100 GeV 
(right panel; $H_0$ is fixed to $67\rm \ km\ s^{-1}\ Mpc^{-1}$ in the calculation of the intrinsic EGB spectrum).

\subsection{Fitting results}

We use the modeled EGB spectrum to fit the $>$10 GeV observed data.
$H_0$ and $\Omega_{\rm m}$ are set to free, and the other parameters are fixed to those in \citet{Ajello} and in \citet{Razzaque}.
The Markov Chain Monte Carlo (MCMC) technique is used to perform our fitting. More details of our
MCMC method can be found in \citet{Yan}.

\begin{figure*}
 \centering
     \includegraphics[scale=0.523]{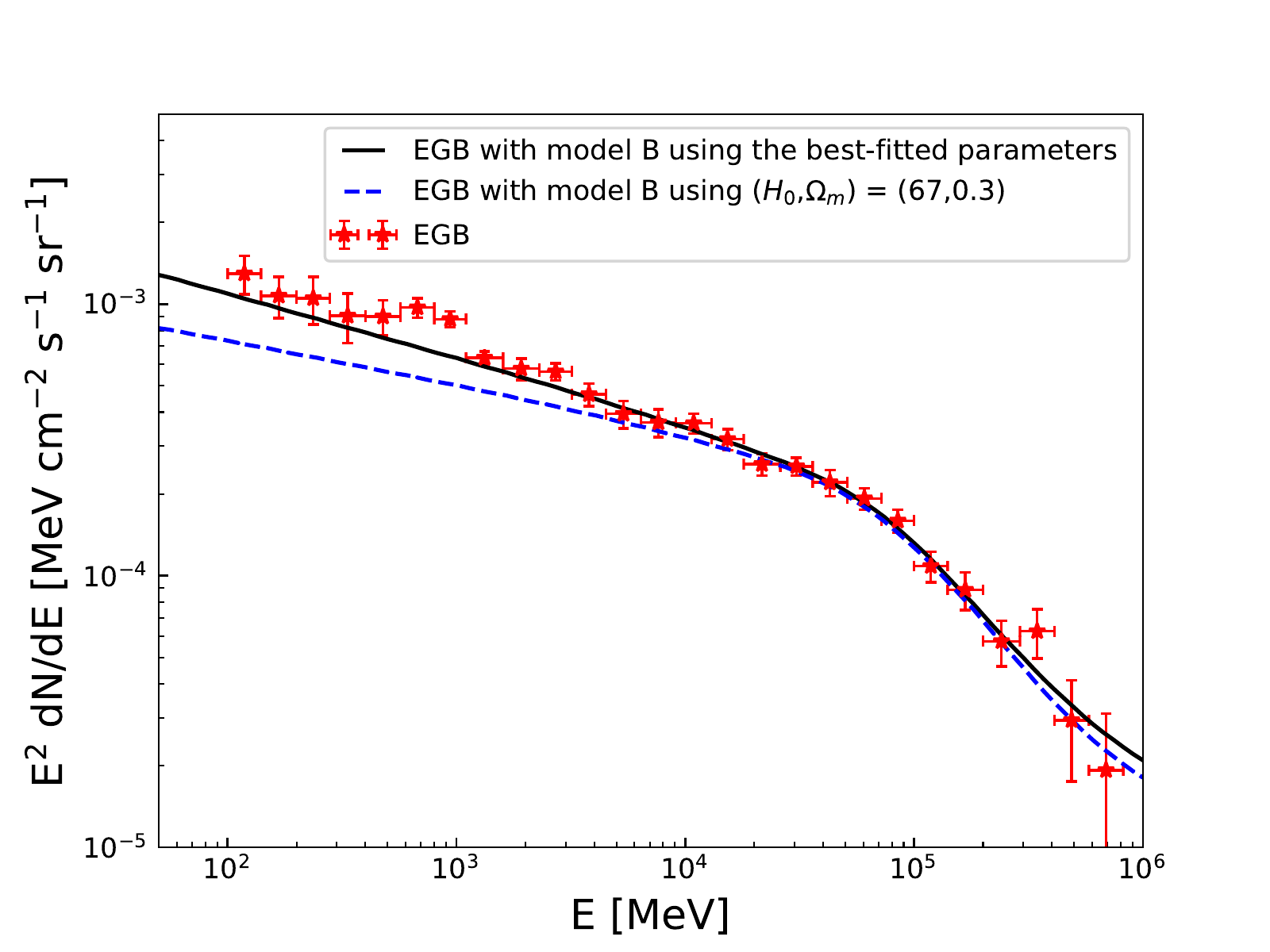}
     \includegraphics[scale=0.263]{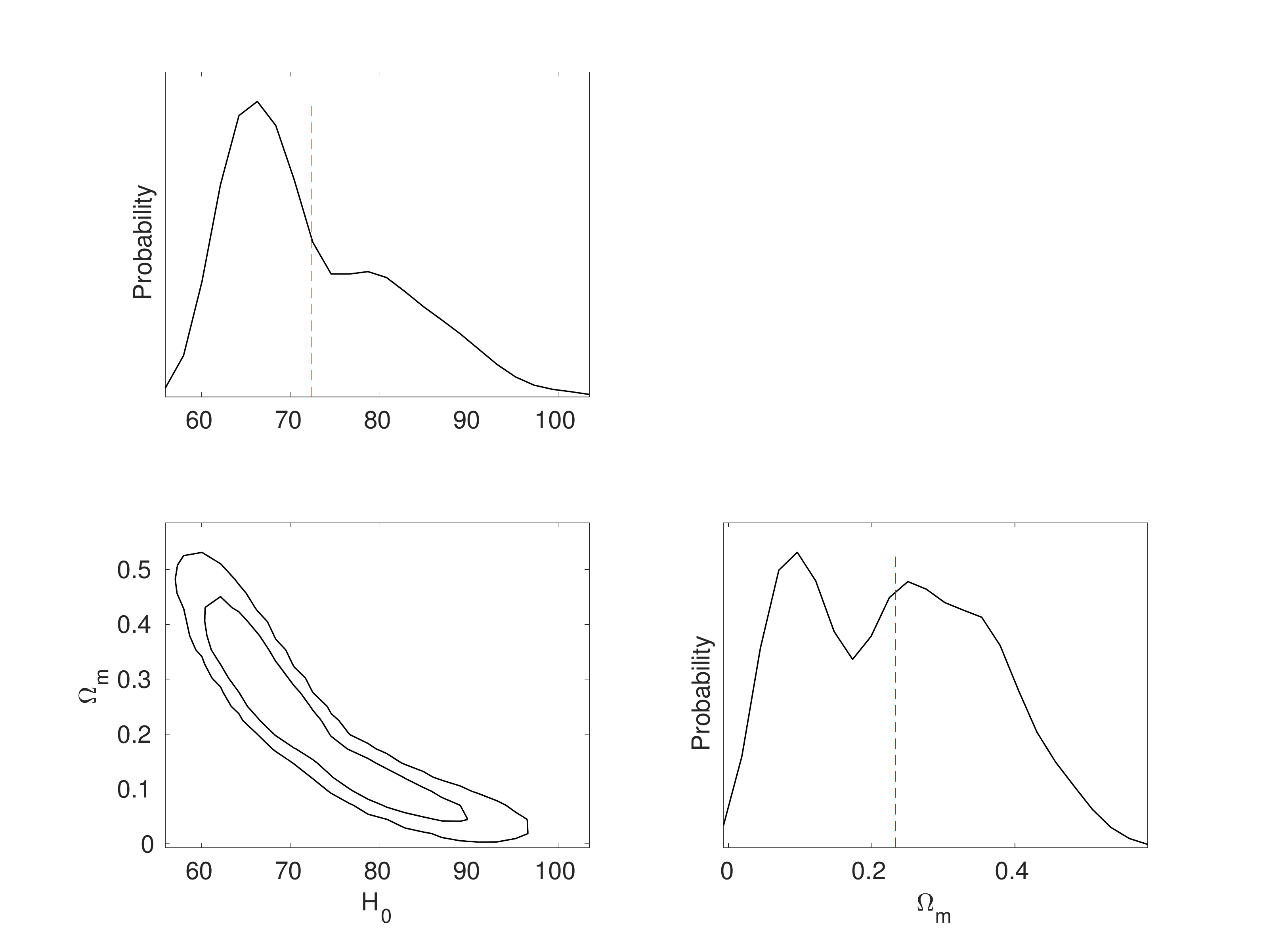}
\caption{Fitting results with the EBL {\it Model B}. {\it Left}: best-fitting to the EGB spectrum above 10 GeV (solid line), 
              and the result calculated with $H_0=67\rm \ km\ s^{-1}\ Mpc^{-1}$ and $\Omega_{\rm m}=0.3$ (dashed line).
               {\it Right}: 1D marginalized probability distribution and 2D confidence contours of the parameters, and the dashed line represents the mean of the parameter. } 
\label{fitB}
\end{figure*}

Fig.~\ref{fitB} shows the best-fitting results with EBL {\it Model B}.
We obtain  $H_0=72^{+10}_{-9}\rm \ km\ s^{-1}\ Mpc^{-1}$ and $\Omega_{\rm m}=0.23^{+0.14}_{-0.13}$ \footnote{We here report the posterior probability means for the parameters.}.
In the fitting, $H_0$ is anti-correlated with $\Omega_{\rm m}$ (see the 2D confidence contours of the parameters in the right panel), 
which is consistent with the result obtained by using the EBL model of \citet{Finke10} in \citet{Dom19}.
We note that the calculated EGB spectrum below 5 GeV is more sensitive to $H_0$ and $\Omega_{\rm m}$ (see the solid and dashed lines in the left panel of Fig.~\ref{fitB}).
This effect is brought by the LF.

\begin{figure*}
 \centering
     \includegraphics[scale=0.523]{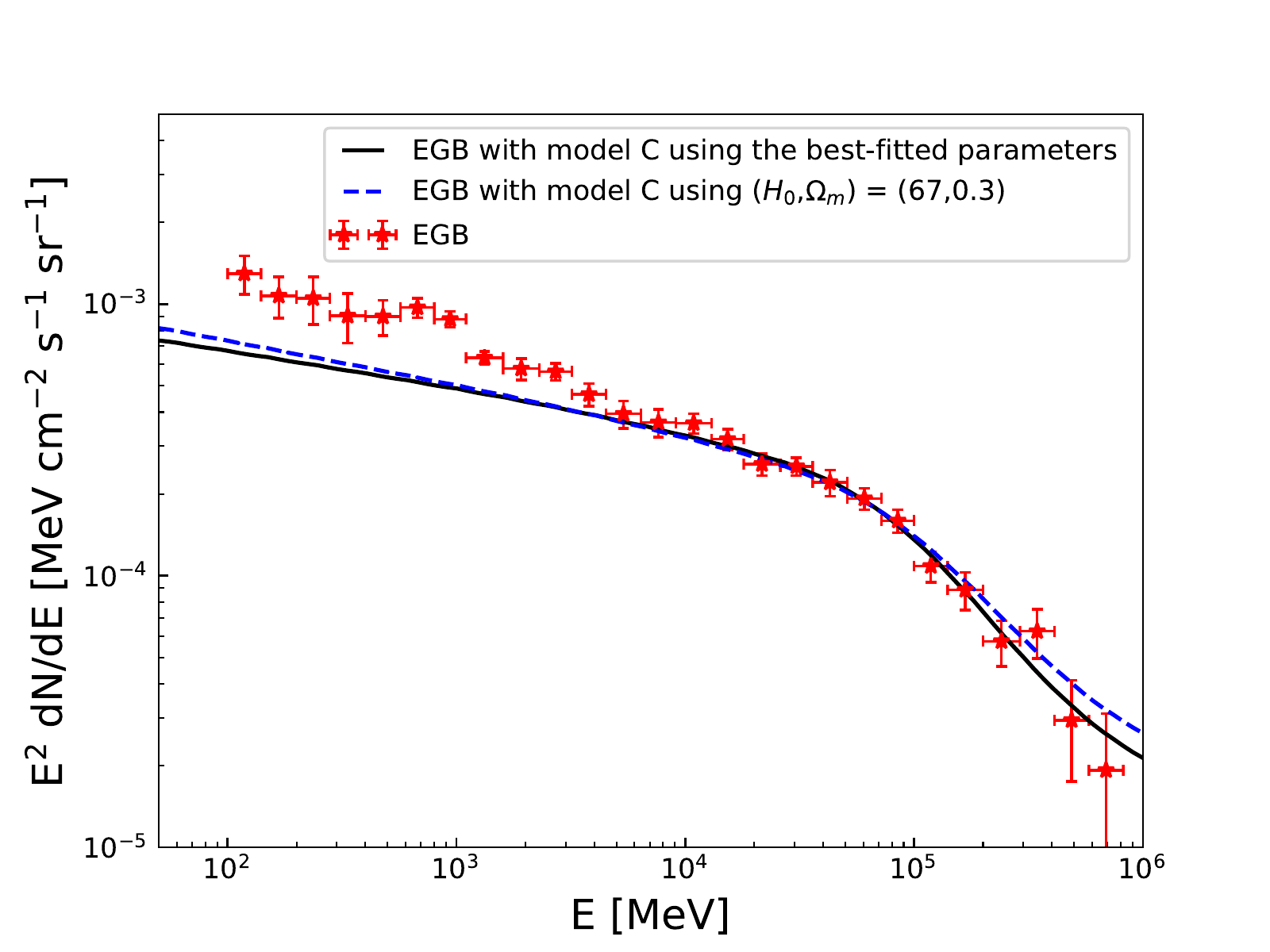}
     \includegraphics[scale=0.263]{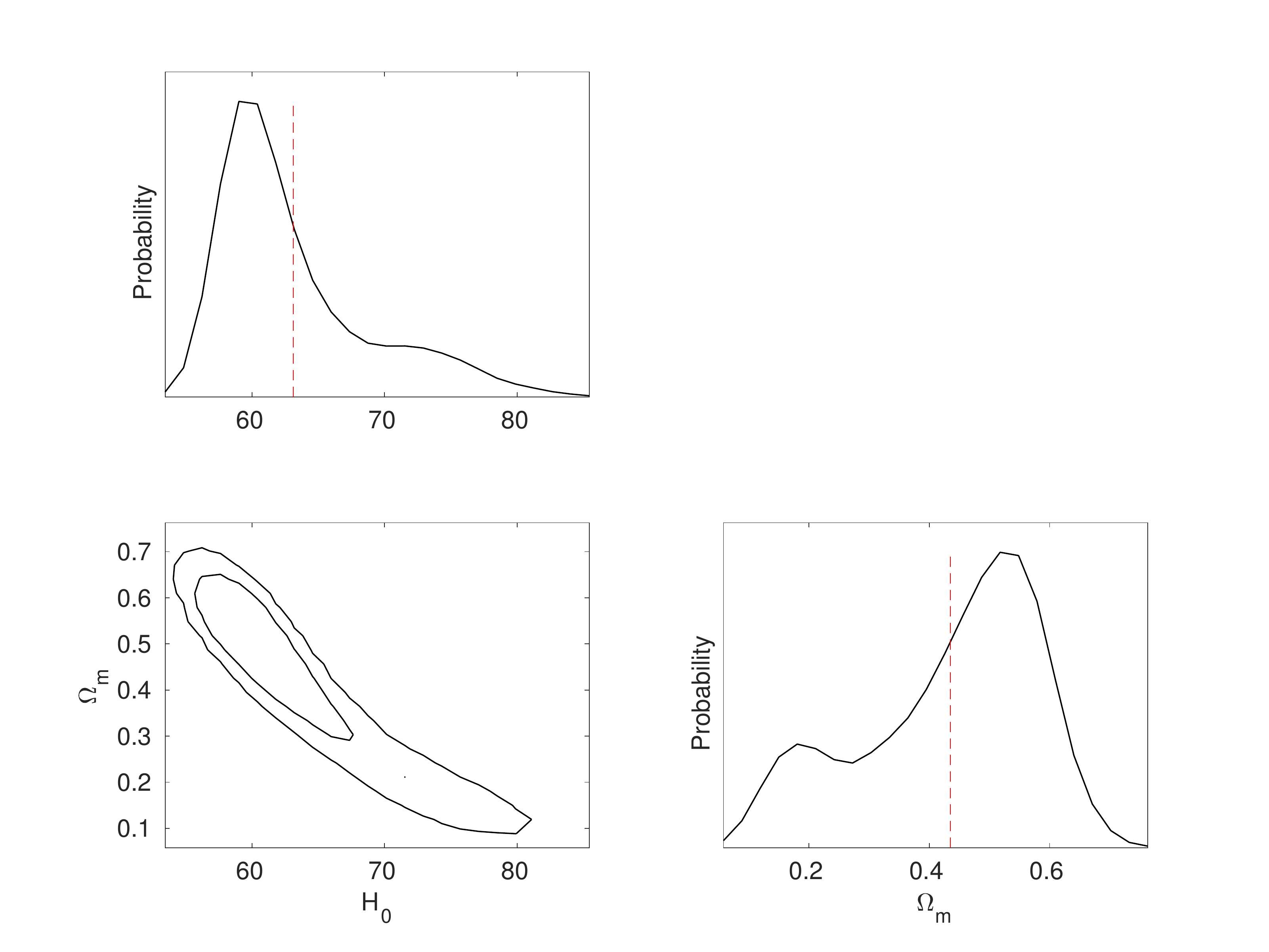}
\caption{Same as Fig.~\ref{fitB}, but with the EBL {\it Model C}.} 
\label{fitC}
\end{figure*}

Fig.~\ref{fitC} shows the best-fitting results with EBL {\it Model C}.
We obtain  $H_0=63.1^{+6.2}_{-4.7}\rm \ km\ s^{-1}\ Mpc^{-1}$ and $\Omega_{\rm m}=0.44^{+0.13}_{-0.19}$.
The uncertainties on $H_0$ are at the 9\% level.
Again, there is a strong degeneracy between $H_0$ and $\Omega_{\rm m}$ in this model.
The EGB spectrum calculated with $H_0=67\rm \ km\ s^{-1}\ Mpc^{-1}$ and $\Omega_{\rm m}=0.3$ is almost same with the best-fitting EGB spectrum.

\begin{figure*}
 \centering
     \includegraphics[scale=0.4]{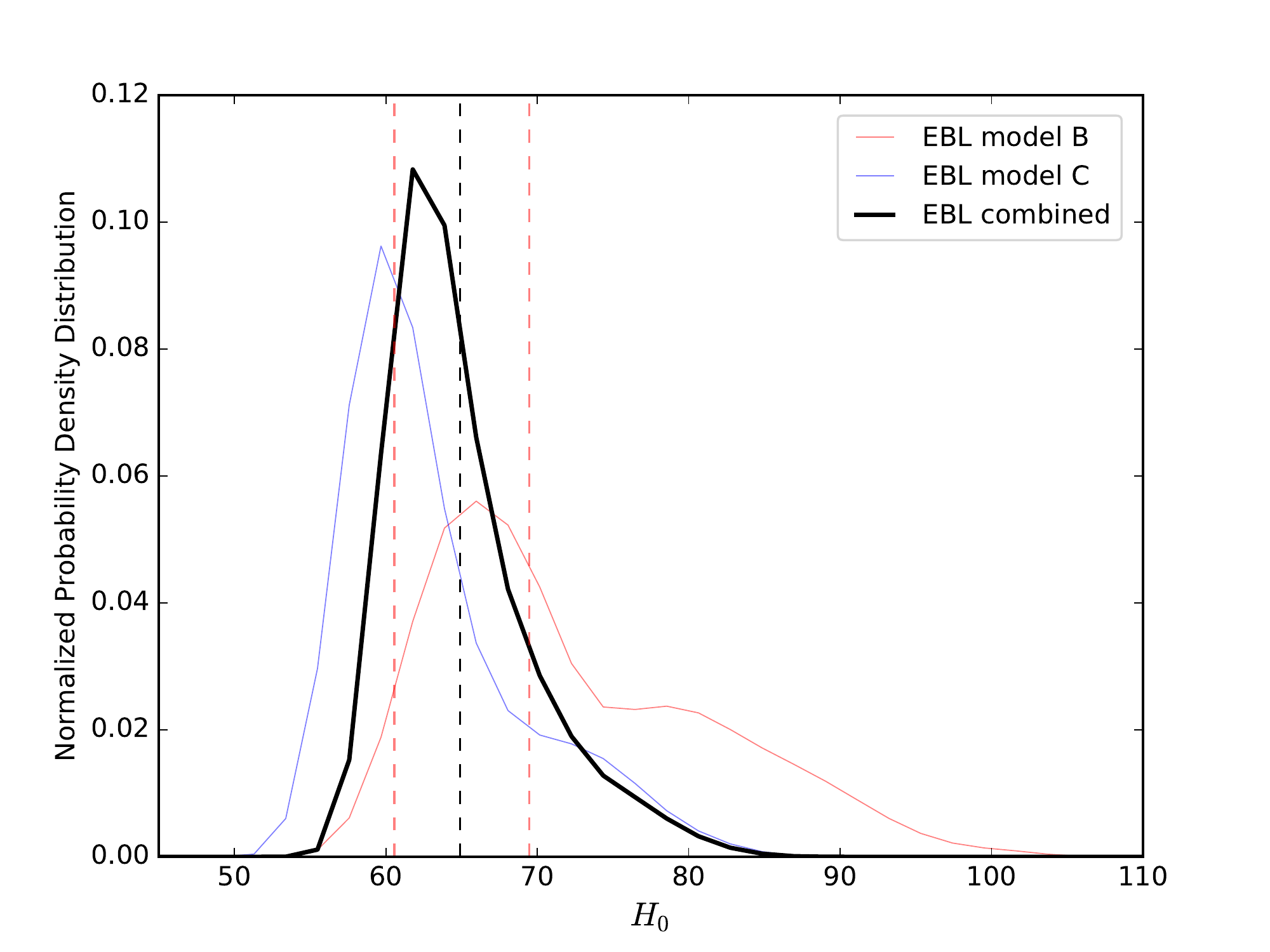}
     \includegraphics[scale=0.4]{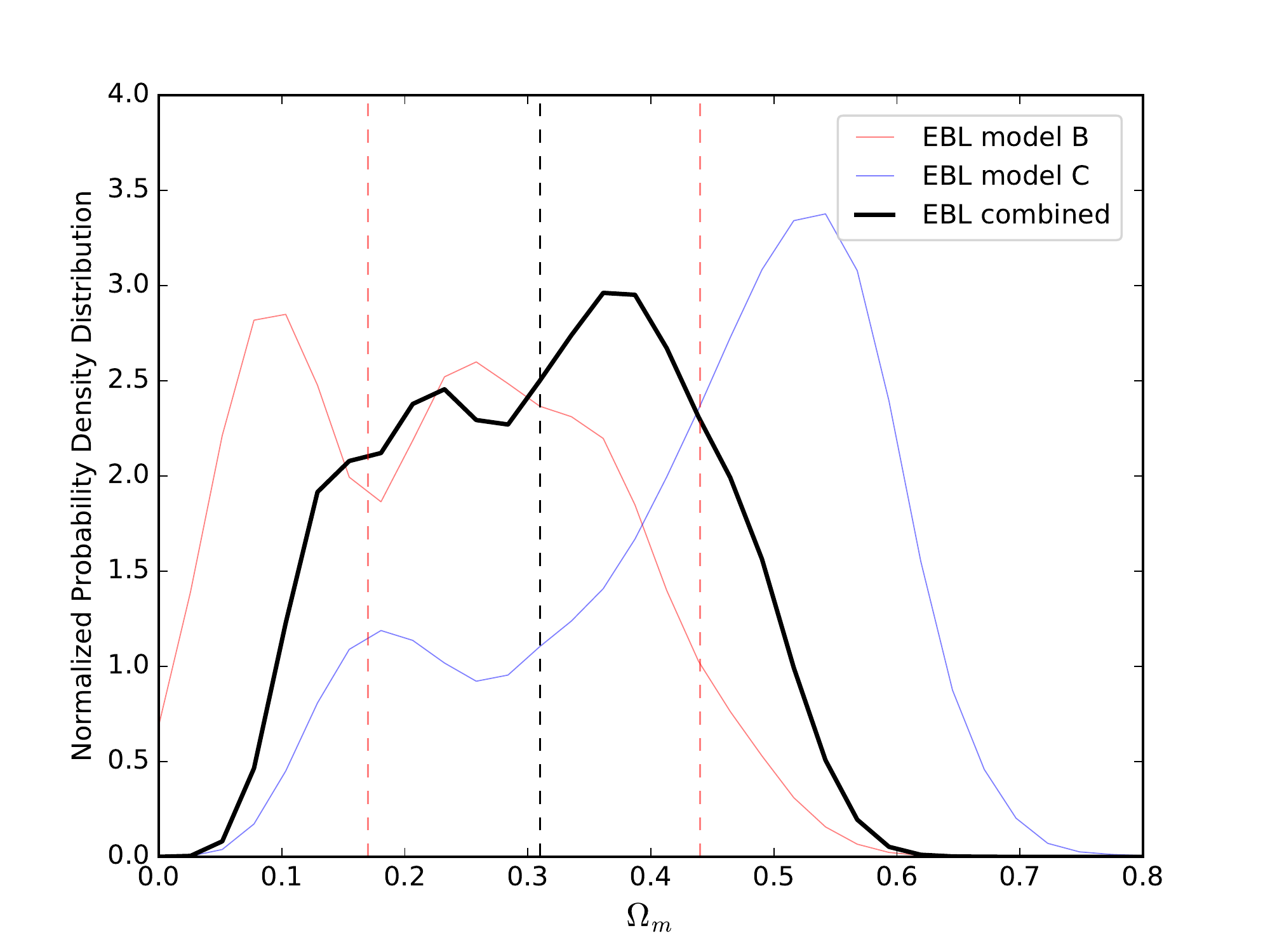}
\caption{Combined results for $H_0$ and $\Omega_{\rm m}$ (black solid line). The blue and orange solid lines are the results with the EBL {\it Model C} and {\it Model B}, respectively.
               Vertical dashed lines are the mean (black) and 1$\sigma$ limits (red) of the combined results. The two models are considered equally likely.} 
\label{comb}
\end{figure*}

Supposing that the two EBL models are equally possible, we derived the combined results in Fig.~\ref{comb} of $H_0=64.9^{+4.6}_{-4.3}\rm \ km\ s^{-1}\ Mpc^{-1}$ and $\Omega_{\rm m}=0.31^{+0.13}_{-0.14}$.

\section{Discussion and Conclusions}

We simultaneously constrain $H_0$ and $\Omega_{\rm m}$ via fitting the $>10$ GeV EGB spectrum.
Two EBL models are adopted to investigate their impacts on the constraints.
The EBL {\it Model B} in \citet{Razzaque} leads to $H_0=72^{+10}_{-9}\rm \ km\ s^{-1}\ Mpc^{-1}$ and $\Omega_{\rm m}=0.23^{+0.14}_{-0.13}$, 
and the EBL {\it Model C} in \citet{Razzaque} leads to $H_0=63.1^{+6.2}_{-4.7}\rm \ km\ s^{-1}\ Mpc^{-1}$ and $\Omega_{\rm m}=0.44^{+0.13}_{-0.19}$.
The constraints obtained by using the two EBL models are consistent.
The combined results are $H_0=64.9^{+4.6}_{-4.3}\rm \ km\ s^{-1}\ Mpc^{-1}$ and $\Omega_{\rm m}=0.31^{+0.13}_{-0.14}$.
Our constraints are mainly given by the blazars below the redshift of 1.5 (see Fig.~\ref{EGB}).

Using the latest $\gamma$-ray attenuation data obtained from $\gamma$-ray spectra of blazars, 
\citet{Dom19} obtained $H_0=71.0^{+2.7}_{-2.6}\rm \ km\ s^{-1}\ Mpc^{-1}$ and $\Omega_{\rm m}=0.21\pm0.06$ with the EBL model of \citet{Finke10},
and $H_0=65.0\pm2.9\rm \ km\ s^{-1}\ Mpc^{-1}$ and $\Omega_{\rm m}=0.14\pm0.06$ with the EBL model of \citet{Dom}.
Their combined results are $H_0=68.0^{+4.2}_{-4.1}\rm \ km\ s^{-1}\ Mpc^{-}$ and $\Omega_{\rm m}=0.17^{+0.07}_{-0.08}$.
Our results are in agreement with theirs.

The uncertainties on $H_0$ are comparable with those obtained by \citet{Dom19}.
There is a clear degeneracy between $H_0$ and $\Omega_{\rm m}$ in our calculation.
Measurement of $\Omega_{\rm m}$ using other independent methods would improve the constraint on $H_0$.

We choose the two easily calculated EBL models to examine the uncertainties introduced by the EBL models.
Actually, these two models belong to the same methodology, i.e., the
physically motivated model.
These two models use the same assumption for SFR, and only differ in IMFs.
Different assumptions for SFR may introduce extra uncertainties on $H_0$.
In addition, we cannot examine the uncertainties introduced by different methodologies of building EBL models \citep[e.g.,][]{Dom19}.
The uncertainties in our results mainly come from EBL models.
Therefore, we may underestimate the uncertainties in our results.

Currently, the values of $H_0$ measured from type Ia
supernovae and from cosmic microwave
background radiation (CMB) are discrepant at $3\sigma$ \citep{Riess}.
Alternative methods of measuring the Hubble constant, like the method presented here, is helpful to understand this discrepancy.

\section*{Acknowledgements}
We thank the referee for the constructive comments. We acknowledge financial supports from the National
Natural Science Foundation of China (NSFC-11703094, NSFC-U1738124, NSFC-11803081, NSFC-11573060, and
NSFC-11661161010) and the joint foundation of Department of Science and Technology of Yunnan Province and Yunnan University [2018FY001(-003)].
The work of D. H. Yan is also supported by the CAS
``Light of West China'' Program and Youth
Innovation Promotion Association.
\bibliography{H0}

\begin{thebibliography}{}
\expandafter\ifx\csname natexlab\endcsname\relax\def\natexlab#1{#1}\fi

\bibitem[Abdollahi et al.(2018)]{FermiSci}Abdollahi, S., Ackermann, M., Ajello, M., et al. 2018, Science,
362, 1031

\bibitem[Ackermann et al.(2015)]{Ackermann}Ackermann, M., Ajello, M., Albert, A., et al. 2015, ApJ, 799, 86

\bibitem[Ackermann et al.(2016)]{Ackermann16}Ackermann, M., Ajello, M., Albert, A., et al. 2016, PhRvL, 116, 151105

\bibitem[Ajello et al.(2015)]{Ajello}Ajello, M., Gasparrini, D., S\'{a}nchez-Conde, M., et al. 2015, ApJL, 800, L27

\bibitem[Baldry \& Glazebrook(2003)]{Baldry}Baldry, I. K., \& Glazebrook, K. 2003, ApJ, 593, 258

\bibitem[Barrau et al.(2008)]{Barrau}Barrau, A., Gorecki, A., \& Grain, J. 2008, MNRAS, 389, 919

\bibitem[Biteau \& Williams(2015)]{Biteau}Biteau, J., \& Williams, D. A. 2015, ApJ, 812, 60

\bibitem[Blanch \& Martinez(2005)]{Blanch}Blanch, O., \& Martinez, M. 2005, Astroparticle Physics, 23, 588

\bibitem[Cole et al.(2001)]{Cole}Cole, S., Norberg, P., Baugh, C. M., et al. 2001, MNRAS, 326, 255

\bibitem[Desai et al.(2019)]{Desai}Desai, A., Helgason, K., Ajello, M., et al. 2019, ApJL, 874, L7

\bibitem[Dom\'{i}nguez et al.(2011)]{Dom}Dom\'{i}nguez, A., Primack, J. R., Rosario, D. J., et al. 2011, MNRAS, 410, 2556

\bibitem[Dom\'{i}nguez et al.(2013)]{Dom13a}Dom\'{i}nguez, A., Finke, J., Prada, F., et al., 2013, ApJ, 770, 73

\bibitem[Dom\'{i}nguez \& Prada(2013)]{Dom13}Dom\'{i}nguez, A., \& Prada, F. 2013, ApJL, 771, L34

\bibitem[Dom\'{i}nguez et al.(2019)]{Dom19}Dom\'{i}nguez, A., Wojtak, R., Finke, J., et al., 2019, arXiv: 1903.12097

\bibitem[Finke et al.(2010)]{Finke10}Finke, J. D., Razzaque, S., \& Dermer, C. D. 2010, ApJ, 712, 238



\bibitem[Gould \& Shr\'{e}der(1967)]{Gould}Gould, R. J., \& Shr\'{e}der, G. P. 1967, Phys. Rev., 155, 1404

\bibitem[Hopkins \& Beacom(2006)]{sfr}Hopkins, A. M., \& Beacom, J. F. 2006, ApJ, 651, 142



\bibitem[Mannheim(1996)]{Mann}Mannheim, K. 1996, RvMA, 9, 17

\bibitem[Razzaque et al.(2009)]{Razzaque}Razzaque, S., Dermer, C. D., \& Finke, J. D. 2009, ApJ, 697, 483

\bibitem[Riess et al.(2018)]{Riess}Riess, A. G., Casertano, S., Yuan, W., et al. 2018, ApJ, 861, 126

\bibitem[Salamon et al.(1994)]{Salamon}Salamon, M. H., Stecker, F. W., \& de Jager, O. C. 1994, ApJL,
423, L1

\bibitem[Salpeter(1955)]{Salpeter}Salpeter, E. E. 1955, ApJ, 121, 161

\bibitem[Suyu et al.(2012)]{Suyu}Suyu, S. H., Treu, T., Blandford, R. D., et al. 2012, arXiv:1202.4459


\bibitem[Yan et al.(2013)]{Yan}Yan, D. H., Zhang, L., Yuan, Q., Fan, Z. H., Zeng, H. D., 2013, ApJ, 765, 122


\end{thebibliography}




\end{document}